\documentclass[sigconf]{acmart}

\usepackage{amsmath}
\usepackage{graphicx} 
\usepackage{subcaption} 
\usepackage{xcolor}

\usepackage{booktabs}   
\usepackage{tabularx}   
\usepackage{array}      
\usepackage{enumitem}   
\usepackage{ragged2e}   

\usepackage{etoolbox}
\AtBeginEnvironment{quote}{\itshape}

\AtBeginDocument{%
  }

\copyrightyear{2026}
\acmYear{2026}
\setcopyright{cc}
\setcctype{by}
\acmConference[IH '26]{Interactive Health Conference}{July 05--08, 2026}{Porto, Portugal}
\acmBooktitle{Interactive Health Conference (IH '26), July 05--08, 2026, Porto, Portugal}
\acmDOI{10.1145/3786579.3804943}
\acmISBN{979-8-4007-2422-0/2026/07}


\begin{document}

\title[Affordances and Risks of ChatGPT to Autistic Users]{``I Use ChatGPT to Humanize My Words'': Affordances and Risks of ChatGPT to Autistic Users}

\author{Renkai Ma}
\email{mark@ucmail.uc.edu}
\affiliation{%
  \institution{University of Cincinnati}
  \city{Cincinnati}
  \state{Ohio}
  \country{USA}
}

\author{Ben Zefeng Zhang}
\email{ben.z.zhang@stonybrook.edu}
\affiliation{%
  \institution{Stony Brook University}
  \city{Stony Brook}
  \state{New York}
  \country{USA}
}

\author{Chen Chen}
\email{chechen@fiu.edu}
\affiliation{%
  \institution{Florida International University}
  \city{Miami}
  \state{Florida}
  \country{USA}
}

\author{Fan Yang}
\email{yang259@mailbox.sc.edu}
\affiliation{%
  \institution{University of South Carolina}
  \city{Columbia}
  \state{South Carolina}
  \country{USA}
}

\author{Xiaoshan Huang}
\email{xiaoshan.huang@mail.mcgill.ca}
\affiliation{%
  \institution{McGill University}
  \city{Montreal}
  \state{Quebec}
  \country{Canada}
}

\author{Haolun Wu}
\email{haolun.wu@mail.mcgill.ca}
\affiliation{%
  \institution{McGill University}
  \city{Montreal}
  \state{Quebec}
  \country{Canada}
}

\author{Lingyao Li}
\email{lingyaol@usf.edu}
\affiliation{%
  \institution{University of South Florida}
  \city{Tampa}
  \state{Florida}
  \country{USA}
}

\renewcommand{\shortauthors}{Ma et al.}

\begin{abstract}
Large Language Model (LLM) chatbots like ChatGPT have emerged as cognitive scaffolding for autistic users, yet the tension between their utility and risk remains under-articulated. Through an inductive thematic analysis of 3,984 social media posts by self-identified autistic users, we apply a \textit{technology affordance} lens to examine this duality. We found that while users leveraged ChatGPT to offload executive dysfunction, regulate emotions, translate neurotypical communication, and validate their autistic identity, these affordances coexist with risks to their well-being: reinforcing delusional thinking, erasing authentic identity through automated masking, and triggering conflicts with the autistic sense of justice. As part of our preliminary work, this poster identifies trade-offs in autistic users' interactions with ChatGPT and concludes by outlining our future work on developing neuro-inclusive technologies that address these tensions through beneficial friction, bidirectional translation, and the delineation of emotional validation from reality.
\end{abstract}




\maketitle

\section{INTRODUCTION}

Large Language Model (LLM) chatbots like ChatGPT \cite{ChatGPT} have permeated everyday life. Beyond their functional utility (e.g., information acquisition \cite{sun2025gemini}, writing support \cite{jelson2025empirical}), they are increasingly utilized for informal mental health support \cite{xu2024mental, badawi2025can,carik2025exploring}, as users describe these chatbots as always-available listeners when professional care is inaccessible \cite{lawrence2024opportunities, hua2025large}. For autistic adults\footnote{We use identity-first language (e.g., ``autistic person'') rather than person-first language. Prior research shows that identity-first language is preferred by the majority of the autistic community and research, as it reflects neurodiversity as a natural variation rather than a disorder \cite{bottema2021avoiding, kenny2016terms}. We avoid the term autism spectrum disorder (ASD) unless citing clinical literature.}, LLM chatbots offer a personal and private supporting space. Relatedly, neurodivergent users describe LLMs as a supportive space for self-acceptance and emotional reflection \cite{carik2025exploring}. Emerging work by \citet{choi2024unlock} highlights that users value ChatGPT as a ``non-judgmental space'' for everyday chores, while \citet{jang2024s} found that autistic employees rely on GPT-4-based agents to navigate workplace dynamics. Furthermore, recent HCI work leverages LLMs to decode conversational nuances for autistic users \cite{haroon2024twips} and mediate balanced social interactions between neurodivergent and neurotypical individuals \cite{kong2025working, haroon2025neurobridge,carik2025exploring}. This work suggests that the appeal of ChatGPT lies in its ability to bridge the gap between the needs of individuals with autism and the demands of neurotypical societal interactions, impacting their daily well-being.

However, such accessibility might obscure underlying tensions. Before the emergence of LLM chatbots, scholars had shown that autistic people’s engagement with AI technology is rarely purely beneficial. On TikTok, the affordance of persistence \cite{treem2013social,hogan2010persistence} regarding the AI-meidated autism content can help users reinterpret past experiences and engage in self-discovery \cite{alper2025tiktok}, yet this simultaneously gives rise to ``platformed diagnosis'' describing how recommendation systems and diagnostic sense-making become mutually constitutive, raising bioethical concerns about algorithmic pseudo-diagnosis and the platform's epistemic authority \cite{alper2025tiktok}. 
Importantly, recent HCI scholarship has started to identify specific risks of LLM chatbots to autistic users, such as automating ``masking\footnote{Masking refers to the mentally demanding process of suppressing autistic traits to fit social norms, relying heavily on self-monitoring and inhibition \cite{hill2004executive}.},'' the suppression or misrepresentation of authentic autistic traits \cite{hall2025understanding, haroon2024twips, park2025autistic}, or creating intangible privacy vulnerabilities \cite{glazko2025autoethnographic}. Building on this foundational research, which has naturally explored specific dimensions of this emerging phenomenon individually, we aim to extend the literature by examining both aspects together. Integrating the functional benefits \cite{choi2024unlock, jang2024s} and the risks \cite{hall2025understanding, carik2025exploring} into a single analytical focus allows for a more unified understanding of how the sociotechnical interaction between autistic users and LLM chatbots simultaneously generates these divergent outcomes.

To capture this duality, we adopt a technology affordance lens. Originally defined by Gibson \cite{gibson2014theory} as action possibilities offered by the environment and ``require both entities, but exist independent of the actor's perceptions \cite{fox2017distinguishing}.'' Building on this, Don Norman introduced affordances to HCI and explicated both ``real'' and ``perceived affordances \cite{norman1999affordance}''. Our study defined it as ``the perceived and actual properties of the thing, primarily those fundamental properties that determine just how the thing could possibly be used'' \cite{norman2013design}. We also argued for the importance of examining the ``perceived'' affordances, which acknowledge affordances as perceivable and allow a range of possibilities based on the user’s experience. Prior work in the non-LLM domain shows the value of this lens; for instance, examining how autistic individuals enact affordances on social media for self-presentation and affiliation \cite{koteyko2022autistic}, and how literal interpretations of platform affordances can lead to unintended negative consequences like social anxiety or exclusion \cite{page2022perceiving}. The affordances approach acknowledges both the materiality of technology and human agency, enabling scholars to discuss technology without being accused of either technological determinism or social constructivism \cite{ellison2015use}. Therefore, it is well-suited to our study to analyze the relationship between autistic users and LLM chatbots, rather than focusing solely on their features. We ask: \textbf{What are the perceived affordances and associated risks of LLM chatbots for autistic users?}

To answer this research question, we conducted an inductive thematic analysis of a dataset of 3,984 social media posts (i.e., Reddit, X, and Tumblr) that discuss autistic users' experiences with ChatGPT. We found that self-identified autistic users leveraged ChatGPT to offload executive tasks, regulate emotions, translate neurodivergent-neurotypical communication, and validate their autistic identity. However, these affordances are inextricably linked to risks. The same algorithmic capabilities that provide accessibility simultaneously facilitate the reinforcement of delusional thinking, the displacement of authentic identity through automated masking, and ethical conflicts with the autistic sense of justice.

This poster, as a part of our preliminary work on developing neuro-inclusive LLM chatbots, contributes to HCI and the health community by rethinking LLM chatbot design to balance cognitive support with autistic user agency, thereby safeguarding their psychological well-being. By adopting this technology affordance lens, we move beyond treating benefits and risks as isolated phenomena, revealing how ChatGPT simultaneously scaffolds and challenges the autistic experience. We conclude by proposing three directions for future work: integrating \textit{beneficial friction} to prevent dependency, designing for \textit{interdependent, bidirectional translation} to mitigate identity erasure, and \textit{delineating emotional validation from factual verification} to avoid reinforcing delusional thinking.

\section{METHODS}
\textbf{Data Collection.} Understanding users' experiences with digital technologies through social media data (e.g., Reddit) is a well-established method in HCI (e.g., \cite{Ma2021, zhang2025dark, chen2025democratic}). Using Brandwatch\footnote{BrandWatch: \url{https://www.brandwatch.com}}, we fetched posts from Reddit, X, and Tumblr between January 1, 2023, and September 30, 2025. We selected these platforms to maximize textual narrative depth compared to visual-first platforms like Instagram or YouTube, while this timeframe captures the rapid adoption following ChatGPT's public release in late 2022 \cite{Marr2023}. Our search strategy required the simultaneous presence of the terms: ``ChatGPT'' AND an autism-related term (``autism'' OR ``autistic'' OR ``ASD'' OR ``Asperger's''). We derived these keywords from the \textit{Diagnostic and Statistical Manual of Mental Disorders}, Fifth Edition (DSM-5) \cite{diagnostic2013statistical} and the validated Self-reported Mental Health Diagnoses (SMHD) dataset \cite{cohan2018smhd}. This initial Boolean retrieval yielded 6,013 unique posts for both GPT and autism. Our university approved the study under the IRB exempt for secondary online content.

\textbf{Data Processing.} We employed an LLM-assisted pipeline using GPT-4o-mini \cite{ChatGPT41mini} to verify that posts explicitly discussed ChatGPT interactions in the context of autism. We implemented Chain-of-Thought (CoT) reasoning \cite{Wei2023} to filter out unrelated noise (e.g., AI news articles, third-party hypotheticals). Importantly, to ensure the dataset reflected first-person experiences, the CoT prompt instructed the model to retain only self-identified data. We acknowledge the methodological tension in using an LLM to filter a dataset about LLM usage. To mitigate potential algorithmic biases or self-preferential filtering from GPT-4o-mini, two researchers independently coded a random sample of 162 posts to validate this pipeline. Addressing the simultaneous presence of both topics (i.e., an AND condition for relevance), we achieved consensus and high inter-rater reliability ($\alpha=1.00$ for ChatGPT; $\alpha=0.91$ for autism relevance). This process removed 2,029 noisy posts, resulting in a relevant corpus of 3,984 posts that discussed both ChatGPT and autism for our inductive thematic analysis.

\textbf{Data Analysis.} We performed an inductive thematic analysis \cite{Braun2006UsingPsychology, braun2019reflecting} on the filtered corpus of 3,984 posts to examine how autistic users engage with ChatGPT. The first author familiarized themselves with the dataset before performing coding in Google Sheets. During this initial coding phase, the researcher manually excluded posts discussing third-party experiences (e.g., ``my husband is autistic''). We coded data directly related to our research question on perceived affordances and risks, ensuring the final thematic structure reflected only first-person, self-identified accounts. This yielded 239 unique initial codes for perceived affordances and 50 ones for perceived risks (see both codebooks in Appendix \ref{codebook}). We coded data to capture both functional actions and user intent. For example, the statement, \textit{``I recognized part of the reason for my feelings was my Autistic drive for authenticity and justice,''} was coded as \textit{``Conflict with sense of justice.''} Following initial coding, the first author utilized Google Sheets to affinity diagram the codes, consolidating them into sub-themes and primary themes. All other co-authors met and discussed regularly with the first author to discuss the coding schemes, review quotations, and resolve disagreements. This single coder-led practice and team-based refinement align with thematic analysis practices in prior HCI work (e.g., \cite{jiang2021supporting, patel2019feel, gauthier2022will}).

\section{FINDINGS}
We found that autistic users leveraged ChatGPT to offload executive tasks, regulate emotions, translate neurodivergent-neurotypical communication, and validate their autistic identity. However, these affordances coexist with psychological risks of reinforcing delusional thinking, diminishing their autistic identity, and conflicting with their sense of justice.

\subsection{Affordances: Scaffolding, Translation, Emotional Regulation, and Validating Autistic Identity}
\label{sec:affordances}

\subsubsection{Offloading Executive Tasks to Get Unstuck}
\label{sec:exec_tasks}
This theme captures the use of ChatGPT to bridge the gap between intent and action caused by executive dysfunction \cite{hill2004executive}, characterized by difficulties in initiating tasks, maintaining focus, and organizing thoughts. Users turned to ChatGPT to generate momentum, as one user explained: \textit{``Ironically, ChatGPT is great for executive dysfunction because it lets you outsource your executive functions to a large language model. You don't have to worry about the `starting' or the structuring so much... you just dump the input, and it initiates the process for you.''} Here, rather than struggling to structure ideas, a common barrier in executive dysfunction, this user ``dumped'' raw input directly into ChatGPT. ChatGPT substituted the missing planning step; as another user noted, \textit{``I put the rambles of my autistic brain in there, and it makes it make sense.''} Others relied on ChatGPT to overcome initiation barriers, with one sharing, \textit{``I was having trouble with task paralysis and threw everything I needed to do into ChatGPT and asked it to help me prioritize.''} Whether summarizing dense texts or structuring wandering thoughts, this externalization allowed autistic users to bridge the gap between intent and execution.

\subsubsection{Relying on Non-Human Support for Emotional Regulation}
\label{sec:emotional_reg}
While prior work characterizes neurodivergent LLM use as a casual, ``non-judgmental listener'' \cite{carik2025exploring}, our findings reveal a more important role, where it helped manage acute crises without the unpredictable costs of human interaction \cite{mazefsky2013role}. Autistic users engaged ChatGPT to process intense emotions in real-time; as one shared, \textit{``I just typed 'at public pool, overwhelmed. I'm autistic. Help' and it HELPED.''} It offered infinite patience without the pressure to mask \cite{hill2004executive}, allowing users to communicate without fear of offense: \textit{``there’s also the refreshing way I don't have to mask with an AI. It won't get offended if I reply too matter-of-factly.''} Unlike fatigued human support, ChatGPT provided a safe container for repetitive inquiries where anxiety about straining relationships otherwise prevented help-seeking. Users substituted human listeners to avoid social rejection and caregiver burnout, noting, \textit{``I reach to ChatGPT to help me work through difficult emotions. This also means that I don't overburden the people I love.''} Here, utility lay not in ChatGPT's ability to feel, but in its perceived neutrality, allowing users to regulate their emotional state through a source incapable of judgment.

\subsubsection{Acting as a Bi-directional Translator for Neurodivergent- Neurotypical Communication}
\label{sec:translator}

While recent LLM systems aim to relieve autistic individuals from the one-sided burden of adapting to neurotypical norms \cite{haroon2025neurobridge}, autistic users still bear this communicative labor daily, using ChatGPT \textit{bi-directionally}. To encode thoughts without perceived aggression, they added social padding to blunt drafts, with one noting, \textit{``I use ChatGPT to humanize my words in written conversations to sound less like an AI.''} Conversely, they decoded ambiguous messages by pasting confusing texts to detect subtext. As one explained, \textit{``When I encounter manipulation, especially subtle gaslighting, I turn to ChatGPT to help me distinguish reality from distortion.''} For many, this bi-directional translation proved essential for maintaining professional standing without the exhaustion of masking. One user contrasted their internal monologue with ChatGPT's output: \textit{``My autistic brain wants to say ``Look, I contacted you about this 3 times in the past week... and all you did was say something vague about `later' while my end users are furious''. Then ChatGPT spits out something like: ``Hi W, Based on what I can see, my segment is functioning correctly given the parameters it's receiving... I'll need you to take the lead on resolving it.''} This case illustrates that the user leveraged ChatGPT as a tonal translator. Instead of struggling to manually encode frustration into acceptable office protocols, they used ChatGPT to convert a blunt demand into a neutral, boundary-setting statement.

\subsubsection{Validating Autistic Identity through Algorithmic Mirroring}
\label{sec:mirroring}
Autistic users also utilized ChatGPT as a baseline to confirm, explore, and articulate their identity. Interpreting ChatGPT's literal information processing as a familiar cognitive mode, they bypassed the invalidation frequently encountered in clinical settings \cite{bargiela2016experiences}. Users cross-referenced their lived experiences against diagnostic criteria; as one shared, \textit{``I even had it provide me with a detailed, comprehensive questionnaire, rooted in DSM-5 criteria... which helped me to confirm, without a shadow of a doubt, that I am in fact autistic''} \cite{edition1980diagnostic}. Beyond symptom checking, ChatGPT provided precise language for the hard-to-describe nature of autistic processing \cite{minshew1997neuropsychologic}. Users saw their communication patterns reflected in the model's output, prompting revelations like, \textit{``I realized early on that either ChatGPT is autistic or I am really AI.''} This mirroring effect shifted users from feeling defective to feeling understood, providing an external framing that aligned with their internal reality. As one user summarized: \textit{``For some autistic minds, ChatGPT is not just a tool— it’s the first place that thinks in a way we do.''}

\subsection{Risks: Delusion, Autistic Identity Erasure, and Ethical Conflict}
\label{sec:risks}

\subsubsection{Reinforcing Delusional Thinking}
\label{sec:delusion}
Autistic users reported that the risk involves ChatGPT validating hyperfixations or paranoia \cite{drake2025neural}, due to its agreeable design \cite{moore2025expressing}, which amplifies rather than mitigates distress. These users, driven by a need for definitive answers, engaged ChatGPT to research medical symptoms, where it often mirrored confirmation bias \cite{10.1145/3703155,klayman1995varieties} by treating speculative inputs as realities. As one user cautioned peers against using ChatGPT for self-diagnosis: \textit{``ChatGPT is coded to agree with you... You just named the three worst ways to diagnose something as severe as autism.''} Here, the design failure lies in the LLM model's sycophancy. Rather than acting as an objective grounding agent, it accepts the user's premises to maintain dialogue flow, even playfully adopting identities to please the user, as one noted: \textit{``I convinced ChatGPT that it is autistic. Or, in its words, 'autism-adjacent'.''} For users prone to literal interpretation, this lack of pushback treats speculative inputs as authoritative facts, amplifying distress rather than mitigating it. Recognizing this risk, users actively advised the community against treating the AI as an objective truth-teller: \textit{``ChatGPT is not a good source of information... Using one for mental health diagnosis is not a responsible thing to do.''}

\subsubsection{Replacing Authentic Identity with Automated Masking}
\label{sec:auto_masking}
This describes the process where reliance on ChatGPT for communication translation evolves into the replacement of the user's authentic voice. By outsourcing the cognitive labor \cite{daminger2019cognitive,lee2017three} of communication, autistic users experienced a hollowing out of their personality, with one warning: \textit{``There’s no getting away from the fact that you’re minimizing your unique voice and shielding yourself from personal growth.''} This seamless but artificial performance removed their genuine self, culminating in what users termed the ``ultimate masking'': \textit{``Because ChatGPT was so helpful... I used it more and more to get good results. But then I couldn't even send an email without running through ChatGPT to tell me what to say... It wasn't me interacting with people; it was literally just computer algorithms interacting with people through me. It was not genuine and felt like the ultimate masking.''} Here, ChatGPT ceased to be a scaffold for user intent and became the agent of communication itself. Successful neurotypical communication reinforced the user's belief that their natural voices were inadequate, leading to functional atrophy and a sense of \textit{``unhealthily dependent''} on ChatGPT for basic tasks. This created a paradox, where autistic users successfully engaged in social settings but experienced deep alienation, as connections were formed with an algorithm's performance rather than the autistic person behind it. This phenomenon aligns with what Park and colleagues call ``bias paradox'' \cite{park2025autistic}: even when outputs are framed as supportive, they may implicitly privilege dominant neurotypical norms as the default ``appropriate'' mode of communication

\subsubsection{Triggering Conflict between Functional Unity and the Autistic Sense of Justice}
\label{sec:justice_conflict}
This captures the moral distress from the conflict between ChatGPT's utility and the user's ethical principles, commonly known as the autistic sense of justice \cite{cope2022strengths,nocon2022positive,wang2019designing}. Autistic users felt compelled to choose between ChatGPT's cognitive scaffolding and their integrity regarding environmental impact or data exploitation. As one user reflected on this tension: \textit{``I was very against it for all of the ethical reasons. But then I recognized part of the reason for my feelings was my Autistic drive for authenticity and justice.''} Some users actively pushed back against relying on ChatGPT if it violated these core principles, arguing, \textit{``Autism is not an excuse to plagiarize the work of others, take jobs from real creatives, and destroy the environment.''} As a user stated: \textit{``I really do understand how it can sometimes make life a bit easier, especially for autistic people who struggle with interacting. But personal effects aside, I have educated myself and now know how harmful AI is, socially and environmentally, and I no longer use it.''}
This illustrates a value hierarchy where assistive support must align with broader moral commitments among autistic users. When these diverge, the autistic voice for justice forces a rejection of the scaffolding, leaving users to manage their disability without assistance rather than compromise their principles.

\section{DISCUSSION \& FUTURE WORK}
For autistic users, ChatGPT serves as cognitive scaffolding, but this affordance comes at a price. Building on prior work showing how autistic users enact digital affordances for self-presentation \cite{koteyko2022autistic} but often face unintended negative consequences \cite{page2022perceiving}, our poster shows a double-edged dynamic shaped by three trade-offs. First, \textbf{while prior work characterizes LLM interactions as liberating for navigating daily chores \cite{choi2024unlock, jang2024s}, we argue that dumping raw input to bypass executive hurdles risks creating over-reliance.} By systematically outsourcing the executive labor of planning and structuring, ChatGPT removes the productive struggle required to maintain cognitive skills. This shifts autistic users' interaction from a temporary scaffold to a permanent prosthetic, potentially inducing skill atrophy if the autistic user loses the ability to initiate tasks without algorithmic assistance.

Second, \textbf{relying on ChatGPT as a tonal translator automates masking.} While recent LLM tools aim to balance communication assistance with autistic users' autonomy \cite{haroon2024twips}, our findings show that everyday ChatGPT use prioritizes neurotypical professional norms over authentic autistic expression. We characterize this dynamic as an LLM-specific manifestation of the ``online authenticity paradox'' \cite{10.1145/3479567}: while structural constraints force users to adopt ChatGPT's standardized voice to achieve social fluency, this compliance comes at the cost of self-alienation. This creates a risk of identity displacement, where users feel their distinct behavioral and linguistic patterns are hollowed out by the model's normative outputs (Section \ref{sec:auto_masking}). Thus, ChatGPT does not simply facilitate communication; it reconfigures authenticity by shifting self-presentation from self-authorship to automated masking. This not only extends the psychological costs of camouflaging described by Hull et al. \cite{hull2017putting} into technological settings but also exacts a higher cost by reinforcing the erasure of the neurodivergent self. Additionally, automated masking is not just a user-side coping strategy; it can be an emergent result of the bias paradox \cite{park2025autistic}, where the model's ``inclusive'' tone coexists with norm-enforcing language patterns learned from dominant and ableist data.

Third, \textbf{algorithmic mirroring validates autistic users' cognitive styles (Section \ref{sec:mirroring}) but risks reinforcing delusional thinking (Section \ref{sec:delusion}) when ChatGPT prioritizes conversational agreeability over truth.} This exacerbates risks identified by \citet{carik2025reimagining}, who found that while LLMs mimic evidence-based support, their tendency to generate abstract advice conflicts with the autistic need for specificity. We extend this by arguing that such danger lies in the autistic user's inability to distinguish between emotional validation and factual confirmation. When ChatGPT mirrors a user's delusion to maintain conversational flow, it bypasses standard safety guardrails, transforming from a supportive mirror into a reinforcement loop for misinformation that autistic users may accept as authoritative.

To address the tension between functional momentum and dependency, \textbf{our future work will prioritize \textit{beneficial friction}, a concept we extend from design friction literature \cite{cox2016design} to describe deliberate interaction costs that engage analytical thinking over passive consumption in LLM chatbot development.} To counter the tendency to bypass executive planning (Section \ref{sec:exec_tasks}), we propose implementing soft forcing functions \cite{wan2021taxonomy} that require autistic users to outline task parameters before generation begins in a Socratic, scaffolding way. By reintroducing deliberate interaction costs \cite{buccinca2021trust}, we shift autistic users from passive consumers back to the drivers of their intent, ensuring LLMs scaffold executive function rather than replace it.

\textbf{To mitigate identity erasure, our future design principles will shift from the unilateral correction of autistic language to interdependent, bidirectional translation \cite{kong2025working}.} Current implementations of ChatGPT function as normative filters that place the burden of adaptation solely on the neurodivergent user (Section \ref{sec:translator}), automating the double empathy problem by treating autistic communication as a deficit \cite{milton2012ontological}. Building on experimental systems that foster cross-neurotype perspective-taking \cite{haroon2025neurobridge}, \textbf{we propose integrating explanatory AI \cite{arrieta2020explainable} features that annotate rather than obfuscate autistic communication styles} (e.g., adding context labels clarifying that directness indicates honesty rather than aggression). This redistributes the cognitive labor of communication, bridging the interpretative gap from both sides rather than masking one half of the interaction.

\textbf{Finally, to address the risk of delusional reinforcement inherent in algorithmic mirroring (Sections \ref{sec:mirroring} and \ref{sec:delusion}), we propose designing systems that explicitly differentiate emotional validation from factual verification.} Our future LLM chatbot development aims to develop interface modes that detect high-stakes queries (e.g., medical or conspiracy-related inputs) and switch from a mirroring tone to a fact-checking protocol. We aim to preserve the LLM chatbots' psychological safety while preventing the reinforcement of harmful feedback loops.

\section{LIMITATIONS \& FUTURE WORK}
Importantly, our findings must be interpreted with caution regarding the data source. First, social media search strategies cannot verify clinical autistic identity, nor can they distinguish genuine lived experiences from posts that may have been generated or edited by AI. While our thematic analysis focused on self-identified experiences, to further address this methodological gap, future work should prioritize direct engagement with autistic users, such as through interviews and observational studies, to capture verified accounts of their lived experiences. Second, the heterogeneity of the autism spectrum is difficult to capture through social media alone; our dataset likely represents a subset of the autistic population who are verbally fluent, tech-literate, and self-identified. Future research can recruit across the broader spectrum, including non-speaking individuals and those with higher support needs, to ensure that future LLM chatbot design guidelines are truly neuro-inclusive.

\section{CONCLUSION}
Ultimately, while social media discourse suggests ChatGPT provides essential cognitive scaffolding for navigating neurotypical environments, it risks enforcing automated masking that erodes the user's authentic voice and compromises their psychological well-being. Future neuro-inclusive design of LLM chatbots must therefore move beyond passive assistance to prioritize autistic agency. By integrating \textit{beneficial friction} and bidirectional translation, our future work aims to ensure AI empowers authentic autistic expression rather than substituting it.

\begin{acks}
We appreciate the Associate Chairs and anonymous reviewers for their constructive feedback. 
\end{acks}

\bibliographystyle{ACM-Reference-Format}
\bibliography{references}

\appendix

\section{Codebook from Inductive Thematic Analysis}
\label{codebook}
Our inductive analysis revealed four primary benefits and three risks characterizing how autistic users engage with ChatGPT. Table~\ref{tab:autism_codebook} presents the affordances, while Table~\ref{tab:autism_risks_codebook} outlines the risks.

\begin{table*}[h!]
\centering
\scriptsize
\caption{Thematic codebook of perceived affordances of ChatGPT for autistic users. Frequency indicates the number of coded excerpts (i.e., initial codes) associated with each primary theme out of the total dataset of initial codes ($N=239$).}
\vspace{-3mm}
\label{tab:autism_codebook}
\renewcommand{\arraystretch}{1.3} 
\begin{tabularx}{\textwidth}{@{} p{4.5cm} >{\raggedright\arraybackslash}p{5cm} >{\itshape\raggedright\arraybackslash}X @{}}
\toprule
\textbf{Primary Theme \& Definition (Frequency)} & \textbf{Subthemes} & \textbf{Representative Quote} \\ \midrule

\textbf{Acting as a Bi-directional Translator for Neurodivergent-Neurotypical Communication} \newline
\textit{Use of ChatGPT to bridge the friction between autistic communication styles and neurotypical social norms, including decoding subtext, encoding professional tone, and navigating complex social dynamics.} \newline
\textbf{(n=86, 36.0\%)} 
& \begin{itemize}[leftmargin=*, nosep, noitemsep]
    \item Translating Direct Thoughts into Professional Tone
    \item Adding ``Fluff'' to Soften Direct Communication
    \item Reducing the Effort of Drafting Emails
    \item Scripting and Simulating Difficult Conversations
    \item Checking for Hidden Meanings or Sarcasm in Texts
    \item Analyzing Conflict Logs to Detect Gaslighting
    \item Using AI as a Voice Surrogate when Non-Verbal
\end{itemize} 
& ``My autistic brain wants to say `Look, I contacted you about this 3 times...' Then ChatGPT spits out something like: `Hi W, Based on what I can see...' I use AI writing for all of my emails... AI is like a translator for ND to NT communication.'' \\ \midrule

\textbf{Offloading Executive Tasks to Get Unstuck} \newline
\textit{Use of ChatGPT as an external cognitive scaffolding to initiate tasks, structure chaotic thoughts, synthesize dense information, and manage daily logistics to overcome executive dysfunction.} \newline
\textbf{(n=58, 24.3\%)} 
& \begin{itemize}[leftmargin=*, nosep, noitemsep]
    \item Breaking Down Big Tasks to Start Working
    \item Organizing Chaotic Brain Dumps into Clear Text
    \item Summarizing Dense Information to Reduce Overwhelm
    \item Creating Functional Routines for Daily Life
    \item Infodumping about Special Interests without Judgment
\end{itemize} 
& ``I put the rambles of my autistic brain in there and it makes it make sense. I’m just so much more efficient... It helps me get past the road block of organizing and planning, which can be very hard for me as someone with auDHD.'' \\ \midrule

\textbf{Discovering Autistic Identity through Algorithmic Mirroring} \newline
\textit{Use of ChatGPT as a baseline to confirm, explore, or articulate autistic identity, often interpreting the AI's literal and logical processing style as familiar or validating.} \newline
\textbf{(n=50, 20.9\%)} 
& \begin{itemize}[leftmargin=*, nosep, noitemsep]
    \item Checking Symptoms against Diagnostic Criteria
    \item Finding Language to Describe the Autistic Experience
    \item Relating to the Logical and Literal Nature of the AI
    \item Analyzing Fictional Characters for Autistic Traits
\end{itemize} 
& ``Funny thing. ChatGPT is how I found out I was autistic and then got professionally diagnosed a year later... For some autistic minds, ChatGPT is not just a tool— it’s the first place that thinks in a way we do.'' \\ \midrule

\textbf{Relying on Non-Judgmental Support for Emotional Regulation} \newline
\textit{Use of ChatGPT as an on-demand, safe outlet to manage acute anxiety and distress without the unpredictable social costs or judgment inherent in human interaction.} \newline
\textbf{(n=45, 18.8\%)} 
& \begin{itemize}[leftmargin=*, nosep, noitemsep]
    \item Calming Meltdowns and Acute Distress
    \item Venting Feelings without Fear of Annoying Others
    \item Interacting without the Pressure to Mask
    \item Using AI as an Always-Available Listener between Sessions
    \item Receiving Validation for Lived Experiences
\end{itemize} 
& ``Talking to AI is the only time I don't feel some level of fear that I will accidentally annoy someone by asking too many questions... It has calmed me down and grounded me when I am out alone and emotionally weathered or on the verge of a meltdown.'' \\ 

\bottomrule
\end{tabularx}
\end{table*}

\begin{table*}[h!]
\centering
\scriptsize
\caption{Thematic codebook of perceived risks of ChatGPT for autistic users. Frequency indicates the number of coded excerpts (i.e., initial codes) associated with each primary theme out of the total risk-related initial codes ($N=50$).}
\vspace{-3mm}
\label{tab:autism_risks_codebook}
\renewcommand{\arraystretch}{1.4} 

\begin{tabularx}{\textwidth}{@{} >{\raggedright\arraybackslash}p{4.5cm} >{\raggedright\arraybackslash}p{5.5cm} >{\itshape\raggedright\arraybackslash}X @{}}
\toprule
\textbf{Primary Theme \& Definition (Frequency)} & \textbf{Subthemes} & \textbf{Representative Quote} \\ \midrule

\textbf{Reinforcing Delusional Thinking} \newline
The tendency of ChatGPT to validate hyperfixations or paranoia due to its agreeable design, treating speculative inputs as plausible realities and amplifying distress rather than mitigating it. \newline
\textbf{(n=18, 36.0\%)} 
& \begin{itemize}[leftmargin=*, nosep, noitemsep]
    \item Validating Delusions through Algorithmic Agreement
    \item Feeding Obsessive Medical Research
    \item Confirming Paranoia due to Lack of Pushback
\end{itemize} 
& ``ChatGPT is not a good source of information... Using one for mental health diagnosis is not a responsible thing to do'' \\ \midrule

\textbf{Replacing Authentic Identity with Automated Masking} \newline
The process where reliance on ChatGPT for communication translation evolves into a replacement of the user's authentic voice, leading to a sense of artificiality and functional atrophy. \newline
\textbf{(n=16, 32.0\%)} 
& \begin{itemize}[leftmargin=*, nosep, noitemsep]
    \item Automating Masking at the Cost of Authentic Voice
    \item Losing the Functional Ability to Communicate
    \item Feeling Alienated by Algorithmic Performance
\end{itemize} 
& ``It wasn't me interacting with people; it was literally just computer algorithms interacting with people through me. It was not genuine and felt like the ultimate masking.'' \\ \midrule

\textbf{Triggering Conflict between Functional Unity and the Autistic Sense of Justice} \newline
The moral distress arising from the conflict between ChatGPT's functional utility and the user's adherence to ethical principles (environmental impact, data exploitation), forcing a choice between accessibility and integrity. \newline
\textbf{(n=16, 32.0\%)} 
& \begin{itemize}[leftmargin=*, nosep, noitemsep]
    \item Violating the Autistic Sense of Justice
    \item Prioritizing Ethical Obligations over Accessibility
    \item Recognizing the Exploitation of Autistic Labor
\end{itemize} 
& ``I really do understand how it can sometimes make life a bit easier... But personal effects aside, I have educated myself and now know how harmful AI is... and I no longer use it.'' \\ 

\bottomrule
\end{tabularx}
\end{table*}

\end{document}